%% file: SysCon2020.tex
\documentclass[10pt,conference]{IEEEtran}
\usepackage{amsmath,array,graphicx}
\usepackage{kantlipsum}
\usepackage{url}
\usepackage{color}
\usepackage{cite}
\usepackage[table,xcdraw]{xcolor}
\usepackage{amsmath,amsfonts,amssymb,amsthm}
\usepackage{mathtools}
\usepackage{threeparttable}
\usepackage{algorithm}
\usepackage[]{algpseudocode}
\usepackage{amssymb}
\usepackage{graphicx}
\usepackage{epsfig}
\usepackage{epstopdf}
\usepackage{subfigure}
\usepackage{stfloats}
\usepackage{multirow}
\usepackage{url}
\usepackage{rotating}
\usepackage{moresize}
\usepackage{amsthm}
\usepackage{amsfonts}
\usepackage{dsfont}
\usepackage{graphicx,changepage}
\usepackage{multirow}
\usepackage[english]{babel}
\usepackage{color}
\usepackage{booktabs}
\usepackage{multirow}
\usepackage{colortbl}
\usepackage{booktabs}
\usepackage{multirow}
\usepackage{colortbl}
\usepackage{graphicx} 
\usepackage[english]{babel}
\usepackage[utf8]{inputenc}
\usepackage{algorithm}
\usepackage{amsmath}
\usepackage{balance} 
\usepackage{flushend}
\usepackage{algpseudocode}
\usepackage{algorithmicx}

\usepackage [english]{babel}
\usepackage [autostyle, english = american]{csquotes}

\algnewcommand\algorithmicforeach{\textbf{for each}}
\algdef{S}[FOR]{ForEach}[1]{\algorithmicforeach\ #1\ \algorithmicdo}
\usepackage[belowskip=-10pt,aboveskip=0pt]{caption}
\begin{document}

\title{Machine Learning-Based Self-Compensating Approximate Computing}
\author{\IEEEauthorblockN{Mahmoud Masadeh, Osman Hasan, and Sofi\`{e}ne Tahar}
\IEEEauthorblockA{Department of Electrical and Computer Engineering, Concordia University, Montreal, Quebec, Canada\\
Email:\{m\_masa, o\_hasan, tahar\}@ece.concordia.ca } }

\maketitle

\input{Abstract}
\input{Introduction}
\input{RelatedWork}
\input{Methodology}
\input{Results}
\vspace{0.7cm} 
\input{Conclusion}
\bibliographystyle{IEEEtran}
\bibliography{main}
\begin{figure}[t!]
\centering
\includegraphics[width=0.99\columnwidth]{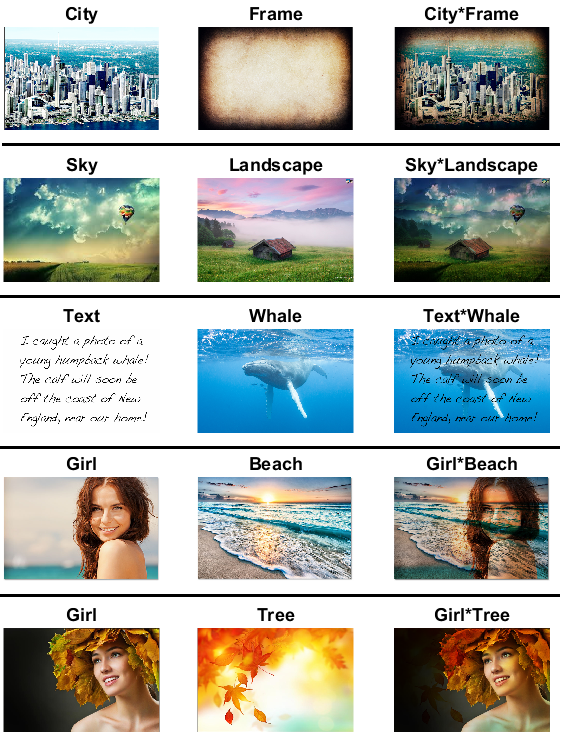} 
\caption{Image Blending Examples for Module Evaluation}
\label{fig:BlendingExamples}
\end{figure}
\end{document}

%% file: Abstract.tex
\vspace{-0.5cm}
\begin{abstract}

Dedicated hardware accelerators are suitable for parallel computational tasks. Moreover, they have the tendency to accept inexact results. These hardware accelerators are extensively used in image processing and computer vision applications, e.g., to process the dense 3-D maps required for self-driving cars. Such error-tolerant hardware accelerators can be designed approximately for reduced power consumption and/or processing time. However, since for some inputs the output errors may reach unacceptable levels, the main challenge is to \textit{enhance the accuracy} of the results of approximate accelerators and keep the error magnitude within an allowed range. Towards this goal, in this paper, we propose a novel machine learning-based self-compensating approximate accelerators for energy efficient systems. The proposed error \textit{compensation module}, which is integrated within the architecture of approximate hardware accelerators, efficiently reduces the accumulated error at its output. It utilizes \textit{lightweight supervised machine learning techniques, i.e., decision tree}, to capture input dependency of the error. We consider image blending application in multiplication mode to demonstrate a practical application of self-compensating approximate computing. Simulation results show that the proposed design of self-compensating approximate accelerator can achieve about 9\% accuracy enhancement, with negligible overhead in other performance measures, i.e., power, area, delay and energy.

 

\end{abstract}

%% file: Introduction.tex
\vspace{-0.2cm}
\section{Introduction} \label{sec:introduction}

Dedicated hardware accelerators are extensively being advocated to be used in complex heterogeneous system-on-chip to process large data more efficiently than pure software processing \cite{HwAcc}. Moreover, hardware accelerates have a reduced power consumption, reduced latency and increased parallelism. These features make them quite suitable for image and digital signal processing (DSP) applications. Approximate computing (AC) or best-effort computing \cite{AC1} is being adapted as a new design paradigm, in both hardware and software \cite{S1}, for error-resilient applications, due to the increased  benefits of approximation, i.e., simplified circuit design with reduced silicon area, delay and power consumption. Several designs of approximate arithmetic components, i.e., adders \cite{XORFA}, dividers \cite{Div2} and multipliers \cite{MasadehGLS}, have been presented. Such approximate components are integrated to form \textit{approximate hardware accelerators} (AxAcc), which are suitable for error-tolerant computationally intensive applications, e.g., big-data and image processing. These applications can tolerate error due to the following factors \cite{ComputingEfficiency}: 1) the lack of a unique, golden result, where a range of results are equally acceptable, 2) no guarantee or need to find the best solution where good-enough result is sufficient, 3) the input data is noisy with iterative-refinement nature, and 4) a reduced quality is tolerable by perceptual, i.e., visual or hearing human limitations.




The approximation error persists permanently during the entire lifetime of the \textit{approximate hardware accelerators} (AxAcc). Thus, it is necessary to develop techniques that can alleviate approximation error and enhance the accuracy with minimal overhead, when high error cannot be afforded. Thus, it is crucial to tackle this issue at the early design stage and change the architecture of \textit{approximate hardware accelerators} by building a lightweight internal error compensation/recovery module with minimal overhead, i.e., area, power and delay.

Despite the unprecedented power saving and reduced execution time introduced by design approximation, it is still an immature computing paradigm \cite{AxCTesting}, where to the best of our knowledge, a formal model of the impact of approximation on accuracy metric is still missing \cite{AxCSecurity}. However, accuracy performance of approximate designs is highly \textit{input-dependent} \cite{ErrorTR}, where we know relatively little about enhancing the accuracy of approximation in a disciplined manner. In this paper, we propose a novel machine learning (ML)-based self-compensating approximate accelerator, aiming to improve the accuracy of the approximated results. There is no clear relationship between the inputs of approximate accelerators and their errors. Therefore, such accelerators are designed by employing ML-based compensation module, to capture input dependency of error. This leads to a noteworthy reduction in error magnitude, with negligible overhead.


As a proof of concept, we consider \textit{approximate hardware accelerators} with 8-bit approximate array multipliers \cite{MasadehGLS}. Such accelerators have 9 bits of the results being approximated. Also, they utilize full adder (FA) cells, known as approximate mirror adder 5 (\textit{AMA5}) \cite{Vaibhav}, which provides a simplified design with reduced area, power and delay. The challenge is to build an efficient compensation module, which considers the value of the inputs. Thus, machine learning techniques are used to capture such dependency. Finally, we consider an image blending application, where two images are multiplied pixel-by-pixel to demonstrate a practical application of \textit{self-compensating approximate hardware accelerators}.  

The rest of the paper is structured as follows: Section \ref{sec:RelatedWork} introduces the related work. Section \ref{sec:Methodology} explains our proposed methodology to enhance the accuracy of approximate hardware accelerators. The obtained results utilizing image processing are described in Section \ref{sec:Results}. Section \ref{sec:Conclusion} concludes the paper and highlights the future work.

%% file: RelatedWork.tex
\section{Related Work} \label{sec:RelatedWork}

There has been significant work on designing approximate components and accelerators. However, to the best of our knowledge, there are very few works targeting the enhancement of the accuracy of approximate accelerators. While most prior works focus on error prediction, in this paper, we aim to overcome the approximation error through an input-dependent error compensation. 


Authors of \cite{Xu} approximated different designs given as behavioral descriptions based on the expected coarse-grained input data distributions. Then, they used these approximate designs to build an adaptive hardware accelerator based on the applied workload. However, the proposed approximate circuits heavily depend on the training data used during the approximation process, where not all possible workload distributions can be precharacterized. Thus the real workload may differ completely from the training one. Authors of \cite{Marcelo} performed a design-space exploration of state-of-the-art approximate designs, and proposed a flow for designing approximate coarse-grained reconfigurable arrays (CGRAs). Green \cite{Green} and SAGE \cite{Sage} check the output quality of approximate programs through sampling techniques, and use a more accurate configuration if the approximation error is high. However, \cite{Marcelo} -- \cite{Sage} are inadequate for fine-grained input data.


A machine learning-based technique has been proposed in \cite{MasadehDATE2019}, aiming to control the quality of approximate computing through selecting the most suitable approximate design based on the inputs. Nevertheless, this technique is efficient when having a set of approximate designs to select the most suitable among them, which is not always applicable. A fault recovery method utilizing machine learning to ameliorate the effect of permanent faults have been proposed in \cite{Taher}, assuming that the number of unique values of error distance (ED) is very low, i.e., less than 5. However, such assumption is unrealistic, where the value of the ED may range from 1 to $2^n$, based on fault location, where \textit{n} is the number of circuit inputs. Recently, a self-compensating accelerator has been proposed in \cite{MAZAHIR20199} by integrating approximate components with their complementary designs, i.e., having the same error magnitude with opposite polarity. However, obtaining such complementary components is not always guaranteed, e.g., the approximate multiplier based on \textit{AMA5}, which is utilized in this work does not have a complementary design. Moreover, the approximate design and its complementary design may have different characteristics, i.e., area, power, delay and energy.

Aiming to avoid the overhead of adapting the design and improving its accuracy, in this paper, we investigate a novel ML-based approach to build an input-dependent \textit{compensation module} for approximate accelerators. The proposed approach relies on the high error rate (ER) of the approximate accelerator aiming to lower the magnitude of the error distance (ED). Our work is orthogonal to the previous related work, where innovatively we utilize ML-based, i.e., decision tree, model to capture input dependency of error. As a proof of concept, we utilize an approximate hardware accelerator with approximate multipliers based on \textit{AMA5} FAs. 

%% file: Methodology.tex
\section{Methodology} \label{sec:Methodology}

\begin{figure}[t!]
\centering
\includegraphics[width=\columnwidth]{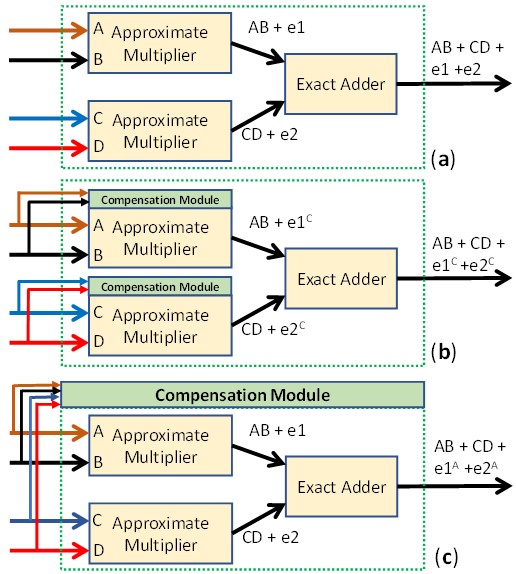}  
\caption{\small{Simplified Architecture for Accelerator of Two Approximate Multipliers, (a) Without Error Compensation, (b) With Error Compensation Module per Approximate Component, (c) With Error Compensation Module per Approximate Accelerator.}}
\label{fig:Accelerator} 
\end{figure}

In self-compensating approximate accelerator, we propose to integrate an input-dependent compensation module in such a way that the accumulative error is reduced. The design of a simplified accelerator with two approximate multipliers is shown in Figure \ref{fig:Accelerator}(a). The magnitude of error $e1$ depends on inputs \textit{A} and \textit{B}, while the magnitude of error $e2$ depends on inputs \textit{C} and \textit{D}. Whereas, $e1$ does not equal $e2$, i.e., $e1$ $\neq$ $e2$, unless $\lbrace A, B \rbrace$ = $\lbrace C,D \rbrace$.  It is important to note that most of the previous work did not consider the input dependency of the approximation error. The final accelerator error is \textit{e}, where $e$ = $e1$ + $e2$. The maximum error is $|e1|$ + $|e2|$. In this paper, without loss of generality, we consider accelerators constructed utilizing 8-bit approximate array multipliers based on \textit{AMA5} FAs with 9-bits of the results being approximated~\cite{MasadehGLS}. However, the proposed methodology is applicable to any approximate accelerator design, e.g., approximate multiply-accumulate units \cite{AxMAC}.


\begin{figure}[t!]
\centering
\includegraphics[width=\columnwidth]{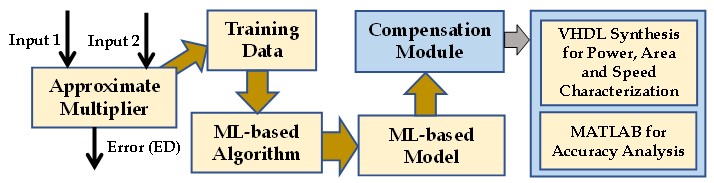} 
\caption{Design Flow for Approximate Accelerator Compensation Module}
\label{fig:SCAxAcc} 
-\vspace{0.2cm}
\end{figure}

The main challenge in the design of self-compensating accelerators is the development of the input-dependent compensation module that has minimal area, delay and power overhead. An overview of the proposed design methodology is given in Figure \ref{fig:SCAxAcc}, where its steps are explained next. The fundamental step in the proposed flow is designing an approximate multiplier, which is the essential building component of the accelerator. Table \ref{tab:ApproxMult} shows the design characteristics of the 8-bit approximate multiplier including its area, delay, power and energy consumption. Moreover, in order to show the benefits of such approximation, the characteristics of the exact multiplier are also shown in Table \ref{tab:ApproxMult}. We evaluate the power, area, delay and energy utilizing the XC6VLX75T FPGA, which belongs to the Virtex-6 family, and the FF484 package. We use Mentor Graphics \textit{Modelsim} \cite{ModelSim}, \textit{Xilinx XPower Analyser} and \textit{Xilinx Integrated Synthesis Environment} (ISE 14.7) tool suite.

\begin{table}[t!]
\centering
\caption{Characteristics of Approximate Accelerator Components, i.e., Approximate Multiplier and Compensation Module}
\label{tab:ApproxMult}
\resizebox{0.99\columnwidth}{!}{%
\begin{tabular}{c|cccccc}
\hline
\multicolumn{1}{|c|}{\textbf{Design}} & \multicolumn{1}{c|}{\textbf{\begin{tabular}[c]{@{}c@{}}Dynamic\\ Power (mW)\end{tabular}}} & \multicolumn{1}{c|}{\textbf{\begin{tabular}[c]{@{}c@{}}Slice\\ LUTs\end{tabular}}} & \multicolumn{1}{c|}{\textbf{\begin{tabular}[c]{@{}c@{}}Occupied\\ Slices\end{tabular}}} & \multicolumn{1}{c|}{\textbf{\begin{tabular}[c]{@{}c@{}}Period\\ (ns)\end{tabular}}} & \multicolumn{1}{c|}{\textbf{\begin{tabular}[c]{@{}c@{}}Frequency\\ (MHz)\end{tabular}}} & \multicolumn{1}{c|}{\textbf{\begin{tabular}[c]{@{}c@{}}Energy\\ (pj)\end{tabular}}} \\ \hline \hline

\multicolumn{1}{|c|}{\textbf{\begin{tabular}[c]{@{}c@{}}Exact\\ Multiplier\end{tabular}}} & \multicolumn{1}{c|}{442} & \multicolumn{1}{c|}{85} & \multicolumn{1}{c|}{33} & \multicolumn{1}{c|}{8.747} & \multicolumn{1}{c|}{114.32} & \multicolumn{1}{c|}{3866.2} \\ \hline

\multicolumn{1}{|c|}{\textbf{\begin{tabular}[c]{@{}c@{}}Approximate\\ Multiplier\end{tabular}}} & \multicolumn{1}{c|}{113} & \multicolumn{1}{c|}{31} & \multicolumn{1}{c|}{11} & \multicolumn{1}{c|}{4.625} & \multicolumn{1}{c|}{216.22} & \multicolumn{1}{c|}{522.6} \\ \hline

\multicolumn{1}{|c|}{\textbf{\begin{tabular}[c]{@{}c@{}}Compensation\\ Module \end{tabular}}} & \multicolumn{1}{c|}{2.79} & \multicolumn{1}{c|}{23} & \multicolumn{1}{c|}{8} & \multicolumn{1}{c|}{2.213} & \multicolumn{1}{c|}{451.88} & \multicolumn{1}{c|}{6.6} \\ \hline

\multicolumn{1}{l}{} & \multicolumn{1}{l}{} & \multicolumn{1}{l}{} & \multicolumn{1}{l}{} & \multicolumn{1}{l}{} & \multicolumn{1}{l}{} & \multicolumn{1}{l}{}
\end{tabular}
}
\vspace{-0.5cm}
\end{table}

\begin{figure}[t!]
\centering
\includegraphics[width=\columnwidth]{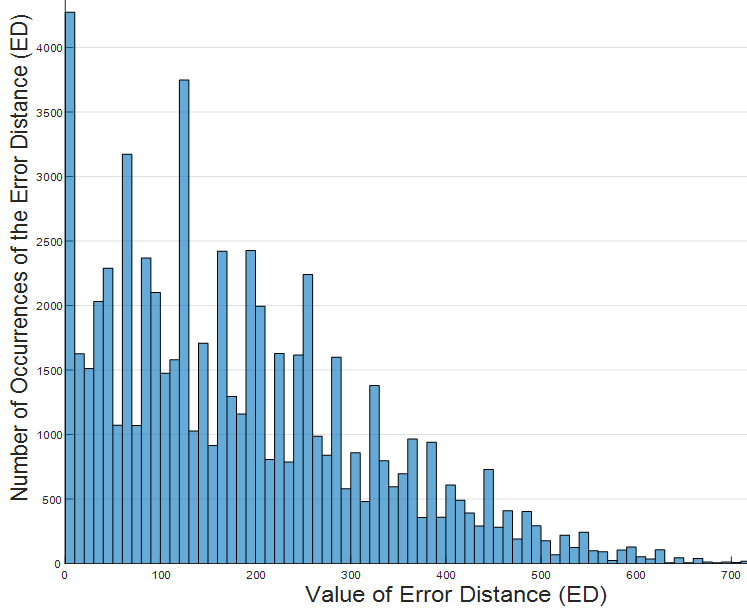} 
\caption{Histogram Distribution of the Error Distance (ED) of the Approximate Multiplier}
\label{fig:HistogramED} 
\end{figure}

Since the magnitude of approximation error is input dependent, we apply an exhaustive simulation by having $2^8=256$ different values for each input. Thus, we have $256*256=65,536$ different input combinations with their associated error distance (ED), which constitute our training data. Figure \ref{fig:HistogramED} shows the histogram distribution for the ED of the approximate multiplier. Accordingly, we can make the following observations regarding the ED:
\begin{itemize}
\item Out of the 65536 possible input combinations, 62420 have inexact results, thus the error rate (ER) is 95.25\%.

\item Approximate computing relies on the principle of fail \textit{small} or fail \textit{rare}. Therefore, high error rate (ER), i.e., 95.25\%, requires having a small value of ED to get an acceptable final result. 

\item Small errors occur more frequently than large errors. For example, we have only 1575 input combinations with ED$>$500, which is about 2.48\% of the erroneous inputs. Considering such extreme values in ED may simplify building the compensation module.

\item Error distance has 176 distinctive values, where the minimum ED is 4, the maximum ED is 756 and the average is 185.
\end{itemize}

Generally, whenever the error occurs for a small fraction of input combinations, i.e., error rate (ER) is low, approximate design with simple error correction, such as adding a constant corrective magnitude, exhibits better performance compared with the exact design. However, our approximate accelerator has an ER of 95.25\%. Therefore, such high ER makes simple error correction inapplicable.


\begin{figure*}[t!]
\centering
\includegraphics[width=0.8\textwidth]{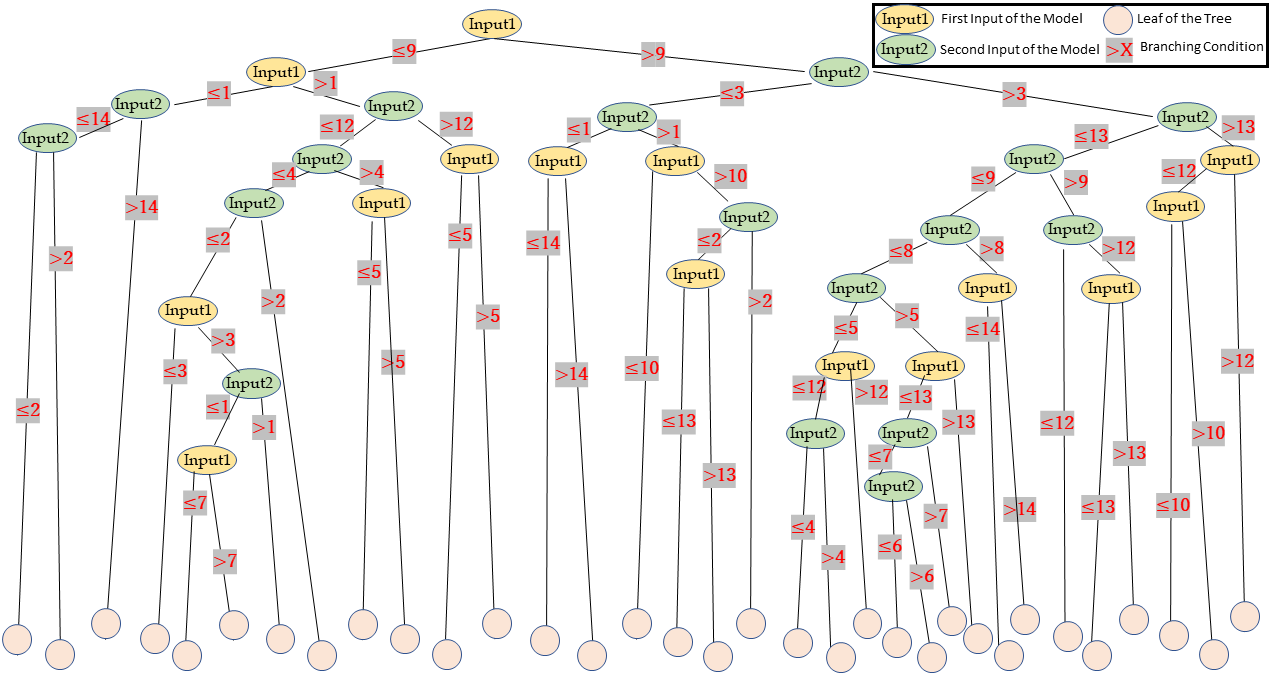} 
\caption{The Structure of the Decision Tree-based Model}
\label{fig:DT} 
\end{figure*}

In order to predict the ED based on the value of the inputs, we use a \textit{lightweight} machine learning-based algorithm, i.e., classification decision tree (DT) based on \textit{C5.0} algorithm~\cite{C50}, given in R~\cite{R} which is a programming and statistical computing language. Decision trees which are fast, memory efficient and have a simple structure, are quite well able to model the non-linear relationship between the inputs and error distance. We notice that the inputs of the approximate design with close magnitudes are associated with a very close ED. Consequently, we quantize the inputs based on their magnitudes into 16 different clusters. Thus, the model has $16*16=256$ different input combinations rather than $256*256=65,536$ which simplifies its internal structure. Figure \ref{fig:DT} shows the structure of the decision tree that we obtained. The leaves of the tree represent the expected values of the ED that should be added to correct the final result, while the internal nodes represent the \textit{conditional decision points} which are the inputs of the model, i.e., the first input (\textit{Input1}) and the second input (\textit{Input2}) of the approximate design. The values associated with the connections between the \textit{conditional decision points} represent the cluster of the inputs, i.e., from 1 to 16. For example, the first branch in Figure \ref{fig:DT} examines the class of \textit{Input1}, then it traces to the left-side if it is $\leq$9 or traces to the right side if the class is $>$9.


To show the effectiveness of the proposed \textit{compensation module}, we perform accuracy evaluation utilizing its implementation in MATLAB. Moreover, we evaluate its power, area, delay and energy. Table \ref{tab:ApproxMult} shows the obtained results, where the power consumption of the module is about 2.8\textit{mW}, which forms about 2.4\% added power to the approximate multiplier. Similarly, the introduced area, delay and energy overhead of the module with respect to the approximate multiplier is about 42.5\%, 32.4\% and 1.2\%, respectively. Such overhead is insignificant when compared to the approximate multiplier where we integrate multiple instances of it within the approximate accelerator.
 
\begin{figure}[t!]
\centering
\includegraphics[width=\columnwidth]{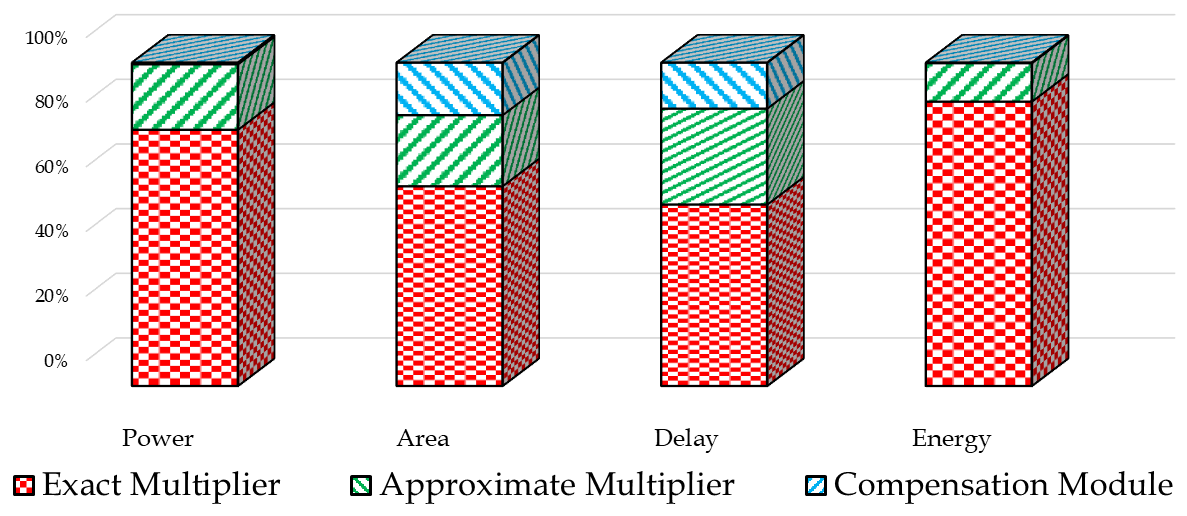} 
\caption{Power, Area, Delay and Energy of Approximate Accelerator Components}
\label{fig:AxC_Module} 
\end{figure}

Figure \ref{fig:AxC_Module} shows a relative representation of the power, area, delay and energy of the approximate multiplier, compensation module as well as the exact multiplier. Despite of the module added overhead, the approximate multiplier with the accompanying module (as shown in Figure \ref{fig:Accelerator}(b)) has a reduction of 73.8\%, 38.1\%, 21.8\% and 86.3\% in the power, area, delay and energy, respectively, compared to the exact multiplier. The error of the approximate multiplier, i.e., $e1$, will be reduced to $e1^C$, which represents $e1$ after being alleviated by the compensation module at the component level, where $e1^C$$<<$$e1$.

Moreover, in order to amortize the overhead of the proposed module, we propose another architectural configuration with a single compensation module for the approximate accelerator, as shown in Figure \ref{fig:Accelerator}.(c), rather than having a dedicated module for each approximate component, as shown in Figure \ref{fig:Accelerator}.(b). Such proposed design is applicable when different data processed at different components have alike values, e.g., adjacent image pixels. Thus, the introduced error is roughly similar. 

In image processing applications, the accelerator processes adjacent image pixels, which usually have close values. Therefore, for image blending in multiplicative mode where the pixels of the two images are multiplied pixel-by-pixel, we propose to divide the image into three segments (colored-components), i.e., red, green and blue. Each colored component is processed on a separate accelerator. For that, the compensation module of the approximate accelerator evaluates the average value of the pixels for each frame colored-component. Based on that, a compensation value is calculated (predicted by decision tree-based model) and then added to all the pixels of the frame colored-component. Thus, the error of the approximate accelerator, i.e., $e1+e2$, will be reduced to $e1^A$ + $e2^A$, based on the error compensation module at the accelerator level. The next section evaluates the accuracy of the implemented \textit{compensating module} that we developed.

%% file: Results.tex
\section{Results and Discussion} \label{sec:Results}

\begin{figure*}[t!]
\centering
\includegraphics[width=0.90\textwidth, height=8cm]{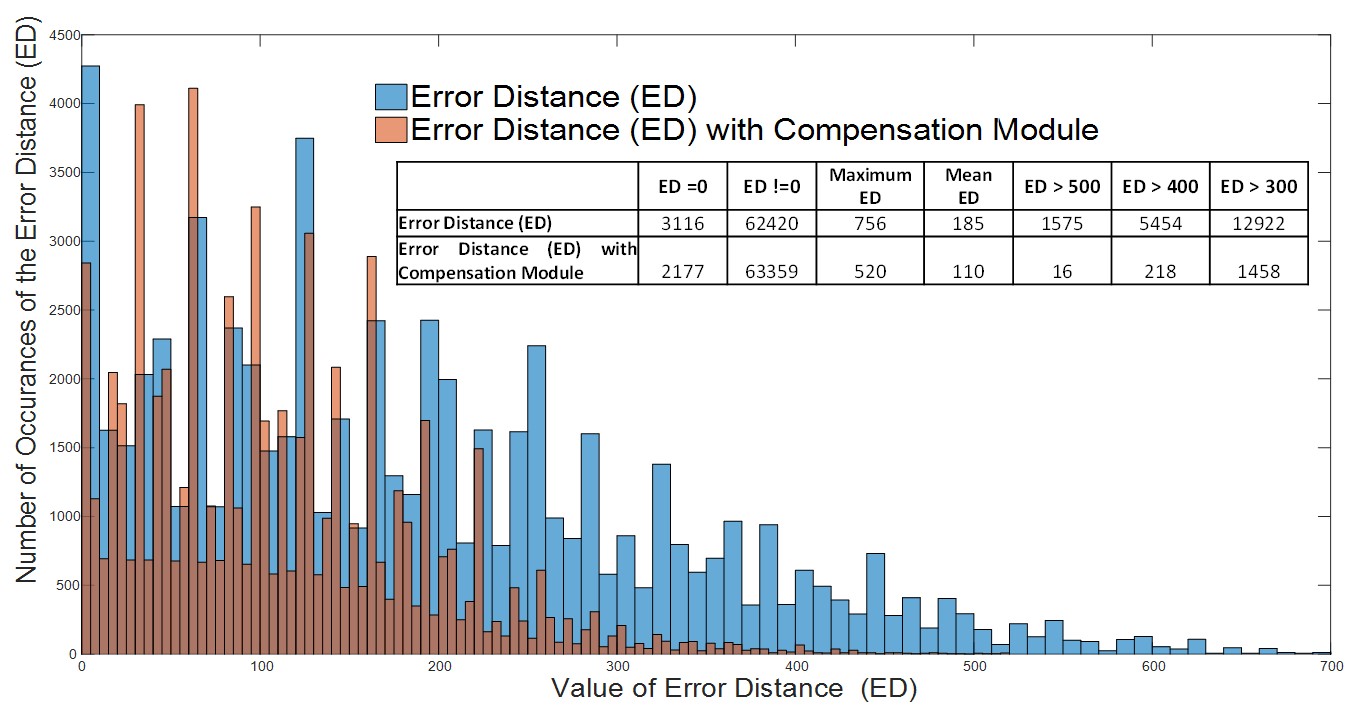} 
\caption{Distribution of Error Distance (ED) of Approximate Multiplier with/without the Error Compensation Module}
\label{fig:HistogramED_Correction} 
\end{figure*}

\begin{figure*}[t!]
\centering
\includegraphics[width=0.90\textwidth]{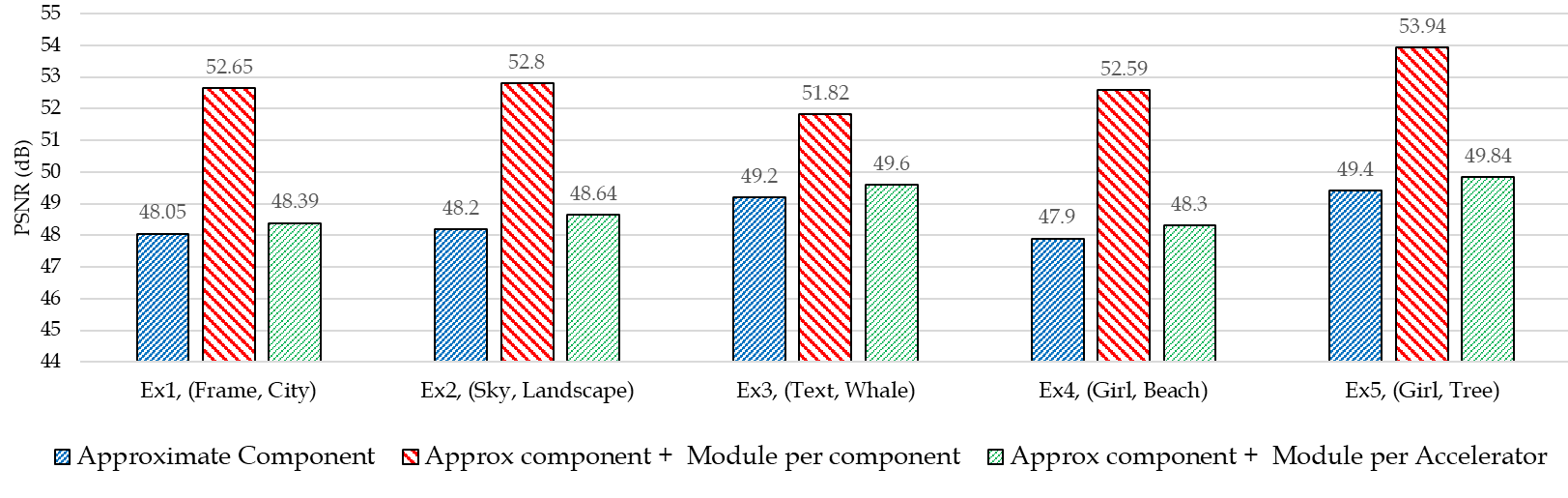} 
\caption{Output Quality (PSNR) of Image Blending, (a) Without Error Compensation, (b) With Error Compensation Module per Approximate Component, (c) With Error Compensation Module per Approximate Accelerator}
\label{fig:BlendingPSNR}
\end{figure*}


This section presents the experimental results obtained by introducing the \textit{compensation module} both at the component and the accelerator level. In order to evaluate the performance of the compensation module, which is shown in Figure \ref{fig:Accelerator}.(b), we perform an exhaustive simulation of the approximate multiplier. Figure \ref{fig:HistogramED_Correction} shows the histogram of the error distance of the approximate multiplier without compensation as well as and the compensated value by integrating the compensation module into the approximate component. Such module will enhance the accuracy of the result, by adding a compensation value based on decision tree-model in order to reduce the final error distance (ED). Clearly, there is a significant reduction in error characteristics, i.e., in both error magnitude and error frequency. 

As summarized in the table shown in Figure \ref{fig:HistogramED_Correction}, the proposed compensation module, reduces the maximum ED of the multiplier from 756 to 520, while the mean ED is decreased from 185 to 110. The number of input combinations with erroneous result, where ED$>$500, is reduced from 1575 input combinations into 16, which is a significant quality improvement. Similarly, the number of input combinations with erroneous result that has an ED$>$400 and ED$>$300 is notably reduced from 5454 to 218, and from 12922 to 1458, respectively. This noteworthy improvement in the quality of results validates the importance of the added compensation module. Moreover, the number of distinctive values of the ED is lowered from 176 to 129. Without the proposed compensation module, the approximate multiplier has 3116 error-free input combinations, i.e., error rate is 95.25\%. However, adding a ML-based compensation module reduces the error-free input combinations into 2177, i.e., error rate is 96.68\%, by erroneously adding a compensation value into error-free result. This is due to \textit{model imperfection}, even though the final accuracy has significant improvement. Similarly, in some cases, the compensation module increase the ED rather than reduce it. Overall, there is a significant reduction in error magnitude and error frequency, where this will enhance the final accuracy of the utilized error-resilient application. 



In order to evaluate the proposed self-compensating approximate accelerators in practical applications, we deployed them in the image blending, where two images are multiplied pixel-by-pixel. The images used in blending and their corresponding accurate results are shown in Figure \ref{fig:BlendingExamples}, where the size of each image is $250$x$400$ pixels. Two configurations of compensation modules are used: 1) a compensation module for each approximate component; and 2) a single compensation module for all approximate components. The Peak-Signal-to-Noise ration (PSNR) of the obtained results are shown in Figure \ref{fig:BlendingPSNR}, which show that the output quality is improved because of error compensation.

As shown in Figure \ref{fig:BlendingPSNR}, all blending examples have an improved quality, i.e., PSNR, whenever the compensation module is used.  Clearly, the improvement in the output quality when the compensation module is incorporated at the component level is higher than the case when the module is used at the accelerator level. The shown results of image blending with error compensation have an enhanced quality, where the increase in the PSNR ranges from 2.6dB to 4.7dB with an average of 4.2bB for the considered examples. Thus, we are able to obtain an average of 9\% improvement in the final quality of image blending application with negligible overhead. Using the compensation module at the accelerator level achieved a lower accuracy enhancement, where the compensation value is evaluated for 100,000 components. Obviously, the accuracy of approximate accelerators can be enhanced by integrating the compensation module at finer granularity level.



%% file: Conclusion.tex
\section{Conclusion} \label{sec:Conclusion}


In this paper, we proposed a novel machine learning-based self-compensating approximate accelerators for enhancing the efficiency of approximate computing applications. In contrast to the state-of-the-art error reduction methodologies, the proposed generic self-compensating methodology has shown an opportunity for error reduction without requiring similar approximate computing elements. The proposed decision tree-based compensation module, illustrated through approximate accelerators, is found to achieve noteworthy enhancement in accuracy performance without compromising the power consumption and speed. This work yields significant new insights into the potential of approximate computing in complex hardware designs, that can lead the designers towards exploiting the problematic error reduction. For future work, we aim to investigate complex accelerators with heterogeneous arithmetic components considering other error metrics rather than ED, as well as other error-tolerant applications. Machine learning based-models, other than decision trees, may be investigated.
